\documentclass{ieeeaccess}
\usepackage{amsmath,amssymb,amsfonts}
\usepackage{algorithm,algorithmic}
\usepackage{graphicx}
\usepackage{textcomp}
\usepackage{hyperref}
\hypersetup{hidelinks}
\usepackage{textcomp} 
\usepackage{balance}
\usepackage{comment}
\usepackage{multirow}
\usepackage{caption}
\usepackage{subcaption}

\usepackage{bm}
\makeatletter
\AtBeginDocument{\DeclareMathVersion{bold}
\SetSymbolFont{operators}{bold}{T1}{times}{b}{n}
\SetSymbolFont{NewLetters}{bold}{T1}{times}{b}{it}
\SetMathAlphabet{\mathrm}{bold}{T1}{times}{b}{n}
\SetMathAlphabet{\mathit}{bold}{T1}{times}{b}{it}
\SetMathAlphabet{\mathbf}{bold}{T1}{times}{b}{n}
\SetMathAlphabet{\mathtt}{bold}{OT1}{pcr}{b}{n}
\SetSymbolFont{symbols}{bold}{OMS}{cmsy}{b}{n}
\renewcommand\boldmath{\@nomath\boldmath\mathversion{bold}}}
\makeatother

\def\BibTeX{{\rm B\kern-.05em{\sc i\kern-.025em b}\kern-.08em
    T\kern-.1667em\lower.7ex\hbox{E}\kern-.125emX}}

\usepackage[
style=ieee
]{biblatex}
\begin{document}
\history{Date of publication xxxx 00, 0000, date of current version xxxx 00, 0000.}
\doi{xxxx}

\title{Beyond the Signal: Medication State Effect on EEG-Based AI models for Parkinson's Disease}
\author{Anna Kurbatskaya\authorrefmark{1}, Fredrik Nilsen Låder\authorrefmark{1}, Andreas Solvang Nese\authorrefmark{1}, Kolbjørn Brønnick\authorrefmark{2,3}, Alvaro Fernandez-Quilez\authorrefmark{1}}

\address[1]{Department of Electrical Engineering and Computer Science, University of Stavanger, Stavanger, 4021, Norway}
\address[2]{Faculty of Health Sciences, University of Stavanger, Stavanger, 4021, Norway}
\address[3]{Stavanger University Hospital, Stavanger, 4068, Norway}

\markboth
{Kurbatskaya \headeretal: Beyond the Signal: Medication State Effect on EEG-Based AI models for Parkinson's Disease}
{Kurbatskaya \headeretal: Beyond the Signal: Medication State Effect on EEG-Based AI models for Parkinson's Disease}

\begin{abstract}
Parkinson’s disease (PD) poses a growing global challenge due to its increasing prevalence, complex pathology, and functional ramifications. Electroencephalography (EEG), when integrated with artificial intelligence (AI), holds promise for monitoring disease progression, identifying sub-phenotypes, and personalizing treatment strategies. However, the effect of medication state on AI model learning and generalization remains poorly understood, potentially limiting EEG-based AI models clinical applicability.
This study evaluates how medication state influences the training and generalization of EEG-based AI models. Paired resting-state EEG recordings were utilized from individuals with PD in both ON- and OFF-medication states. AI models were trained on recordings from each state separately and evaluated on independent test sets representing both ON- and OFF-medication conditions. Model performance was assessed using multiple metrics, with accuracy (ACC) as the primary outcome. Statistical significance was assessed via permutation testing ($p$-values$<0.05$).
Our results reveal that models trained on OFF-medication data exhibited consistent but suboptimal performance across both medication states ($ACC_{\text{OFF-ON}}=55.3\pm8.8$ and $ACC_{\text{OFF-OFF}}=56.2\pm8.7$). In contrast, models trained on ON-medication data demonstrated significantly higher performance on ON-medication recordings ($ACC_{\text{ON-ON}}=80.7\pm7.1$) but significantly reduced generalization to OFF-medication data ($ACC_{\text{ON-OFF}}=76.0\pm7.2$). Notably, models trained on ON-medication data consistently outperformed those trained on OFF-medication data within their respective states ($ACC_{\text{ON-ON}}=80.7\pm7.1$ and $ACC_{\text{OFF-OFF}}=56.2\pm8.7$)
Our findings suggest that  medication state, a hidden confounder, significantly influences the patterns learned by EEG-based AI models. Addressing this challenge is essential to enhance the robustness and clinical utility of EEG-based AI models for PD characterization and management.
\end{abstract}


\maketitle

\section{Introduction}
\label{sec:introduction}
\IEEEPARstart{T}{he} prevalence of Parkinson’s disease (PD) has pronouncedly increased over the last two decades \cite{zhu2024}. 
The current diagnostic pathway relies on a combination of physical examination, clinical evaluation, and patient medical history, requiring significant specialized or trained human resources \cite{postuma2015}. Despite uniform diagnostic criteria, PD manifests with considerable etiological complexity and diversity, translating into varied disease progression patterns among individuals \cite{mulroy2024}. To date, no definitive cure exists but there are effective strategies available to mitigate the disease’s effects \cite{armstrong2020}. Typically, these include a combination of dopamine-based therapies, quantified in terms of levodopa equivalent dose (LED) \cite{jost2023}. As the disease progresses, individuals with PD normally require smaller and more frequent doses, resulting in higher dosages of levodopa \cite{pahwa2009}. The rapidly growing prevalence of the disease and the overarching need for personalized patient management underscores the importance of improving the decision-making process through non-invasive detection and profiling techniques, along with automated data-driven analysis \cite{bloem2021}.

Artificial intelligence (AI) is increasingly being leveraged in medical research, revealing its potential in applying non-invasive techniques such as electroencephalography (EEG) for enhanced understanding of PD and support in clinical decision making \cite{vanneste2018, betrouni2019, waninger2020, khoshnevis2021, yassine2023, sugden2023, suuronen2023, anjum2024}. However, despite the promising developments, the integration of EEG-based AI models and translation of their findings into clinical practice remains limited \cite{frisoni2024}. This translational gap can be partly attributed to a lack of accounting for potential confounding variables, which may limit the studies’ validity and clinical translatability. For instance, biological age and gender have been previously shown to modulate the learning of AI models from EEG recordings \cite{engemann2022,khayretdinova2024}. An oversight of such confounding factors can, for example, cause AI-based models to inadvertently produce biased and disparate outcomes \cite{kurbatskaya2023_1} or lead to the extraction of non-disease-related information, resulting in clinical shortcuts that compromise the reliability and validity of model interpretations.

When recruiting individuals with PD for EEG studies, managing the medication state – whether participants are ON or OFF their medication – is vital to the accurate interpretation of results \cite{stige2024}. Commonly, participants for such studies have clinical manifestations of the disease already present and are undergoing personalized dopamine-based therapies. These medications may considerably alter brain activity as recorded by EEG, thereby potentially impacting study outcomes \cite{babiloni2019, jackson2019}. Specifically, for studies aimed at prodromal PD detection by EEG-based AI models, it is essential to simulate a real-case scenario, which often means having participants in their OFF-medication state \cite{waninger2020}. However, to minimize patients' discomfort and facilitate their cooperation with the study procedure, EEG acquisition is typically conducted in an ON state \cite{vanneste2018}. Furthermore, to enhance the real-world generalizability of models, cohorts from various sites are combined, resulting in the inclusion of participants in different medication states \cite{sugden2023, suuronen2023}. 

The inclusion criteria also vary for other applications of EEG such as cognitive impairment detection \cite{anjum2024}, cognitive profiling \cite{betrouni2019}, identification of PD subphenotypes \cite{yassine2023}, and staging of PD \cite{khoshnevis2021}. The inconsistency in handling the medication state variable can lead to a lack of comparability across studies, making it difficult to draw broad conclusions. Although some studies have examined the impact of medication state on their findings \cite{anjum2020, khare2021, barua2021, aljalal2022}, to the best of our knowledge, no studies have conducted a confounding analysis of the effect of medication state on PD-related outcomes from EEG in the context of AI. Addressing these considerations in study design can lead to more valid and actionable insights, ultimately benefiting clinical practices, patient recruitment, and outcomes in PD management.

Against this backdrop, we hypothesize that the information extracted by EEG-based machine learning (ML) models is significantly altered by the medication state of individuals with PD. To test this hypothesis, we used paired resting-state electroencephalographic (rs-EEG) recordings acquired from the same PD individuals being in different medication states to effectively control for all possible known confounding variables except for medication state. Following, we developed a confounding analysis framework for ML models trained on rs-EEG recordings of individuals presenting clinical manifestations of PD and individuals without PD. Through the framework, we quantified the differences between the performance of the models and assessed the models’ invariance and generalization properties. Given the design of the experiments, any observed differences in models’ performance can be attributed directly to the medication state’s impact, providing a clear assessment of its influence on PD-related predictions of EEG-based ML models.

\section{Data and Methods}
\subsection{Data Cohort}
We used two publicly available datasets from two different research centers: the University of New Mexico, New Mexico \cite{NMdata}and the University of California, San Diego \cite{SDdata}. For the reminder of the article, we refer to them as the New Mexico dataset and the San Diego dataset. The demographic and clinical information for the two datasets is presented in TABLE 1.

\begin{table*}[t]
\centering
\caption{Demographic and clinical characteristics of the two datasets used in this study presented as mean$\pm$standard deviation values. Individuals without Parkinson's disease (non-PD) manifestation had no available data for disease duration, UPDRS III, or LED values at the time of the study.}
\begin{tabular}{l|c|c|c|c|c|c}
& \multicolumn{2}{c|}{New Mexico} & \multicolumn{1}{c}{\multirow{2}{*}{$p$-value}} & \multicolumn{2}{|c|}{San Diego} & \multirow{2}{*}{$p$-value} \\
\cline{1-3} \cline{5-6}
& PD ($n$=25) & \multicolumn{1}{c|}{non-PD ($n$=25)} & \multicolumn{1}{c|}{} & \multicolumn{1}{c|}{PD ($n$=15)} & \multicolumn{1}{c|}{non-PD ($n$=16)} & \\
\cline{1-7}
number of females & 9 (36\%) & \multicolumn{1}{c|}{9 (36\%)} & & 8 (53.3\%) & 9 (56.2\%) & \\ 
\cline{1-7}
\multicolumn{1}{l|}{age, years} & 69.7 (8.7) & \multicolumn{1}{c|}{69.3 (9.6)} & 0.89 & 63.3 (8.2) & 63.5 (9.7) & 0.99 \\ 
\cline{1-7}
MMSE & 28.7 (1.03) & \multicolumn{1}{c|}{28.8 (1.05)} & 0.79 & 28.9 (1.0) & 29.2 (1.1) & 0.99 \\
\cline{1-7}
NAART & 45.9 (9.3) & \multicolumn{1}{c|}{46.8 (7.64)} & 0.72 & 46.1 (6.47) & 49.1 (7.14) & 0.99 \\
\cline{1-7}
disease duration, years & \multicolumn{1}{c|}{5.40 (4.09)} & & & 4.5 (3.5) & & \\
\cline{1-7}
UPDRS III ON & \multicolumn{1}{c|}{23.4 (9.9)} & & & 32.7 (10.42) & & \\
\cline{1-7}
UPDRS III OFF & \multicolumn{1}{c|}{24.8 (8.7)} & & & 39.3 (9.71) & & \\
\cline{1-7}
LED, mg/day & \multicolumn{1}{c|}{685 (452)} & & & 633 (640) & &                         
\end{tabular}
\end{table*}

The New Mexico dataset was collected at the Cognitive Rhythms and Computation Lab at the University of New Mexico, New Mexico, in 2015 \cite{cavanagh2018}. Data gathering was approved by the University of New Mexico Office of the Institutional Review Board, and and all the participants provided written informed consent. Participants were paid \$20/h for their involvement in the study. Individuals with PD ($n$=25) were recruited from the Albuquerque, New Mexico community. Non-PD individuals ($n$=25) were matched by age and gender with the PD individuals. This subset of participants was drawn from a slightly larger sample, with their behavioral data reported in a separate study \cite{cavanagh2017}. There was no difference in education or premorbid intelligence between the two groups. A neurologist performed the Unified Parkinson’s Disease Rating Scale (UPDRS) part III scoring for the PD individuals using videotaped recordings.

The collection of the San Diego dataset was approved by the Institutional Review Board of the University of California, San Diego, and all the participants provided written informed consent \cite{george2013}. PD individuals ($n$=15) were enlisted from Scripps Clinic in La Jolla, California, while non-PD individuals ($n$=16) were either selected from the local community or were spouses of the PD individuals. The two groups were matched by age, gender, handedness, Mini-Mental state Examination (MMSE), and North American Adult Reading Test (NAART) scores for cognitively premorbid matching. Laboratory personnel who had completed online training performed UPDRS III scoring for the PD individuals.

\subsection{Data Acquisition and Preprocessing}
For the New Mexico dataset, one-minute rs-EEG recordings with eyes open were obtained prior to a three-stimulus auditory oddball task. For this study's purpose, only the rs-portion with eyes open of the recordings was extracted. The non-PD individuals participated in a single session, while the PD individuals visited twice, one week apart, either in ON- or OFF-medication state. The OFF-medication state recordings followed a 15-hour overnight withdrawal from subjects’ individual intake of dopaminergic medications. A 64-channel Brain Vision system with sintered Ag/AgCl electrodes was used to record EEG signals within a frequency range of 0.1 to 100 Hz, with a sampling rate of 500 Hz. The CPz and AFz electrodes were used as a reference and ground electrode, respectively. In addition, a vertical electrooculogram was acquired using bipolar auxiliary inputs.

For the San Diego dataset, rs-EEG data were recorded for three minutes. The participants kept their eyes open while focusing on a white cross at the center of a screen. The PD individuals visited the laboratory site twice, both in ON- and OFF-medication state. For the OFF-medication recordings, dopaminergic withdrawal was at least 12 hours. A 32-channel ActiveTwo (Biosemi instrumental system) with a sampling frequency of 512 Hz was used. Two extra electrodes placed on the left and right mastoids were used as reference electrodes. Both vertical and horizontal electrooculograms were obtained using two electrodes placed lateral and below the left eye.

For the preprocessing of the recordings, we followed standardized steps \cite{keil2014}. All the recording files were downloaded in the brain imaging data structure (BIDS) format \cite{pernet2019, appelhoff2019}. All the preprocessing steps were implemented in Python, using the PyPREP \cite{appelhoff2022, bigdely2015} and AutoReject libraries \cite{jas2017} on top of the MNE library \cite{gramfort2013, gramfort2014}. The first step involved robust average re-referencing, adaptive line-noise correlation with a line noise of 60 Hz for both datasets and detection of and interpolation of noisy channels using PyPREP. The recordings were then segmented into two-second epochs and downsampled to 500 Hz. Prior to applying independent component analysis (ICA), the data were high-pass filtered with a cut-off frequency of 1 Hz to remove low-frequency drifts. Additionally, to ensure reliable data for ICA an automated epoch rejection by AutoReject was performed on the filtered epochs. ICA was applied using 15 components and an extended infomax method \cite{lee1999}. Components were annotated using MNE-ICALabel \cite{li2022}. Finally, post-ICA automated epoch rejection using AutoReject was applied. After completing preprocessing, the average number of epochs across recordings was 30 and 87 for the NM and SD dataset, respectively.

Only 29 channels were common across the datasets and were further included in the feature extraction process: AF3, AF4, C3, C4, CP1, CP2, CP5, CP6, Cz, F3, F4, F7, F8, FC1, FC2, FC5, FC6, Fp1, Fp2, Fz, O1, O2, Oz, P3, P4, P7, P8, T7, and T8. Initially, the epoched recordings were low-pass filtered at 30 Hz due to the focus on the lower-frequency bands described below. Power spectral density (PSD) values were computed from each epoch using a multi-taper method \cite{slepian1961}. Subsequently, median PSD values per channel were calculated. Finally, relative band power values were computed for five sub-bands of the frequency spectrum recommended for clinical research: $\delta$ (1-4 Hz), slow-$\theta$ (4-5.5 Hz), fast-$\theta$ or pre-$\alpha$ (5.5-8), $\alpha$ (8-13 Hz), and $\beta$ (13-30 Hz) \cite{babiloni2020, vallat2021}. In total, 145 features were generated per subject, consisting of five relative band power features across 29 channels.

\subsection{Cohort Split}
\begin{figure}
    \centering
    \includegraphics[width=0.9\linewidth]{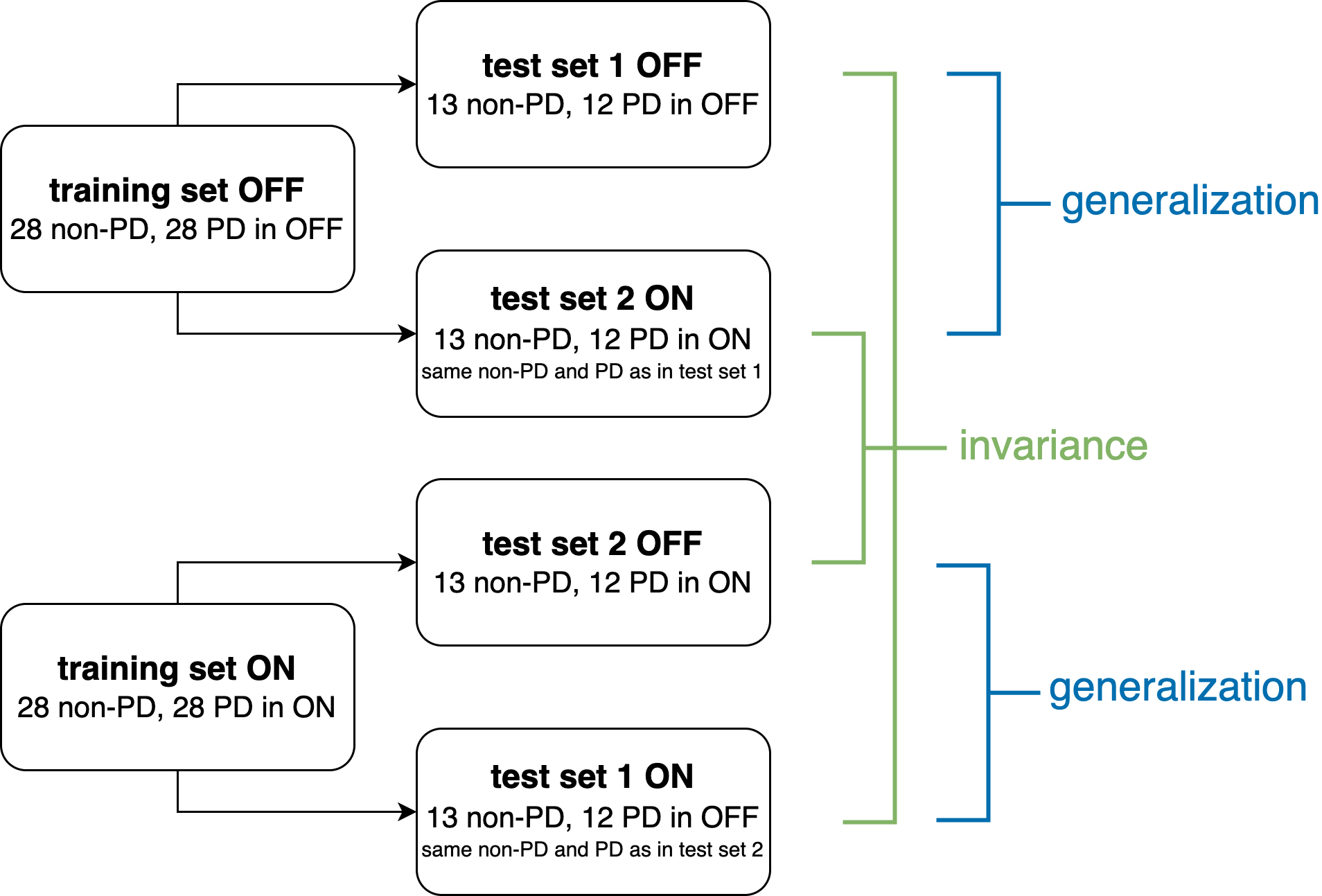}
    \caption{Two data splits were created for training and testing models. The training set OFF contained recordings of individuals with PD in their OFF-medication state, along with non-PD individuals, while the training set ON used recordings from the same PD individuals in their ON-medication state together with the same non-PD individuals as in the training set OFF. For each training set, two test sets were created. Test set 1 consisted of recordings from additional PD individuals in the same medication state and additional non-PD individuals. Test set 2 contained paired recordings from these PD individuals in the alternate medication state, alongside the non-PD individuals from test set 1.}
    \label{fig:split}
\end{figure}

The AI-based models were trained to extract information from EEG correlated with PD, through the task of detecting PD in independent test sets. For this aim, we implemented two distinct data splits into training and test sets presented in Fig. \ref{fig:split}. The first split defined a training set consisting of recordings of PD subjects in their ON-medication state and recordings of non-PD subjects, hereafter referred to as the training set ON. The second split formed another training set including the same PD subjects but in their OFF-medication state, along with the identical non-PD subject recordings from the first split, referred to as the training set OFF.

For each training set, two corresponding test sets were created. The first one included recordings of the remaining PD subjects maintaining the same medication state as its respective training set, along with recordings of the remaining non-PD subjects, further referred to as test set 1. The second test set comprised paired recordings of the same remaining PD subjects but with the opposite medication state, alongside the same non-PD subject recordings as in test set 1, referred to as test set 2. There was one instance of test set 1 and one instance of test set 2 for each training set.

\subsection{Confounding Analysis Framework}
In this analysis, we examine the impact of medication state on the learning process and generalization of the AI models trained on EEG recordings to detect PD. EEG recordings have been shown to covary with the presence of PD, indicating that specific neural oscillatory patterns are affected by PD-related changes in brain activity \cite{vanneste2018, betrouni2019, waninger2020, khoshnevis2021, yassine2023, sugden2023, suuronen2023, anjum2024}. Additionally, the medication state of individuals with PD can significantly affect brain functions in PD, likely introducing an indirect effect on the relationship between EEG and PD-related outcomes \cite{babiloni2019, jackson2019}. This suggests that the medication state can either enhance or obscure the neural dysfunction detectable by EEG, which is typically associated with PD. 

\begin{figure}
     \centering
     \begin{subfigure}[b]{0.4\textwidth}
         \centering
         \includegraphics[width=0.9\linewidth]{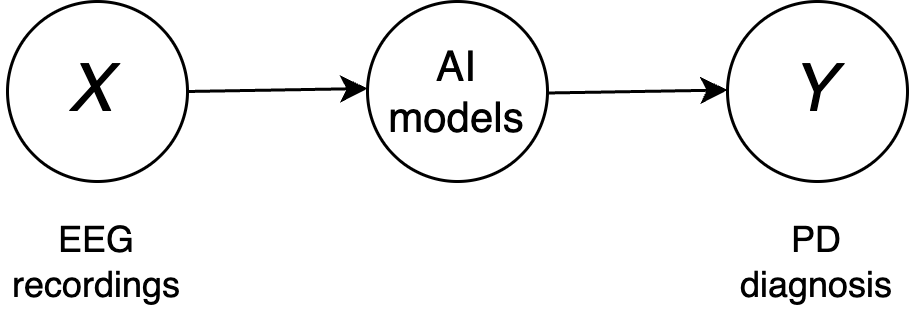}
         \caption{Ideal EEG-based AI prediction of PD diagnosis.}
         \label{fig:causality_1}
     \end{subfigure}
     \hfill
     \begin{subfigure}[b]{0.4\textwidth}
         \centering
         \includegraphics[width=0.9\linewidth]{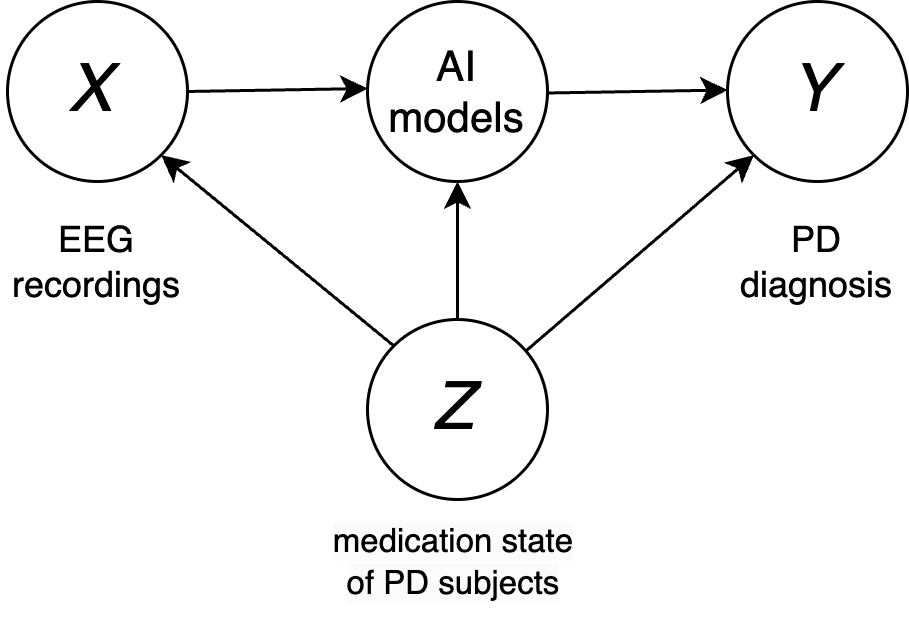}
         \caption{Potential influence of medication state on EEG-based AI prediction of PD diagnosis.}
         \label{fig:causality_2}
     \end{subfigure}
        \caption{A confounding analysis was conducted to assess EEG-based ML models' PD-related predictions while accounting for the potential influence of medication state.}
        \label{fig:causality}
\end{figure}

To explore this, we used rs-EEG recordings as a diagnostic tool of PD, focusing on the relationship $X\rightarrow Y$, where $X$ represents EEG signals, and $Y$ represents PD diagnosis. Specifically, we investigated whether the performance of EEG-based ML models reflects predictions directly related to PD or whether it is instead influenced by the mediator $Z$, representing the medication state. To address this, we propose a confounding analysis framework that assesses the models’ invariance and generalization properties (Fig. \ref{fig:causality}). 

\subsubsection{Assessment of Invariance}
We define an ML-based model’s invariance property as its ability to maintain consistent and valid performance despite irrelevant and undesired variations in the data. In this context, the underlying assumption is that if the information extracted from EEG recordings is related to PD, the medication state should not affect the model’s PD-related outcome. Hence, if the AI models correctly capture disease-related patterns from EEG recordings, they should perform similarly regardless of whether it is trained or tested on recordings of ON- or OFF medication states. In other words, medication should not influence the model’s performance, and the model should be invariant with respect to medication state. 

To test this, we conducted two sets of comparisons. In the first comparison, we trained the models on the training set OFF and tested them on the corresponding test set 1 OFF (Fig. \ref{fig:split}). Following, we trained the models on the training set ON and tested them on the corresponding test set 1 ON. If the AI models capture, exclusively, PD-specific information from rs-EEG recordings, their performance should be consistent across these two training and testing scenarios regardless of whether the data originates from the ON- or OFF-medication state. This comparison evaluates the model's ability to maintain consistent detection capability when both the training and test sets contain medication-related variations. 

In the second comparison, we trained the models on the training set OFF and tested them on the corresponding test set 2 ON, thus introducing medication-related variations only in the test set. Similarly, we trained the models on the training set ON and tested them on the corresponding test set 2 OFF, introducing medication-related variations only in the training set. This approach allowed us to determine if the models could maintain consistent performance across different medication states, regardless of whether the medication-related variations were present in the training or test sets. 

\subsubsection{Assessment of Generalization}
We define an ML-based model’s generalization property as its ability to maintain consistent performance on previously unseen data that were not used during training. In this context, the unseen data were represented by alternating medication states of individuals with PD, which introduced novel information to the models not encountered during the training phase. The underlying assumption is that if the model indeed captures useful information related to PD from the EEG recordings, its performance should remain consistent whether tested on ON- or OFF-medication states. This is because the extracted PD-related information from EEG should be equally relevant for both cases, assuming the model is capturing true disease-related signals rather than medication-induced effects. 

To examine the generalization of the ML-based models, we compared their performance across different test sets (Fig. \ref{fig:split}). For the models trained on the training set OFF, we evaluated their performance when tested on both test set 1 OFF, which matched the medication state during training, and test set 2 ON, which introduced the alternating medication state. Similarly, for the models trained on the training set ON, we tested them on both test set 1 ON, matching the medication state in the training set, and test set 2 OFF, introducing the alternating medication state. This analysis enabled us to assess whether differences in medication state between training and testing impacted the models' performance. 

\begin{table*}[!t]
\centering
\caption{Assessment of invariance: Test results (mean $\pm$ standard deviation (SD)) achieved with a support-vector machine (SVM) model trained and tested on different medication states (ON/OFF). * $p$-values $< 0.05$ were considered statistically significant. Abbreviations: AUC = area under the receiver operating characteristic curve.}
\begin{tabular}{c|c|c|c|c|c|c}
\cline{2-3} \cline{5-6}
& \begin{tabular}[c]{@{}l@{}}train OFF\\ test ON\end{tabular} & \begin{tabular}[c]{@{}l@{}}train ON\\ test ON\end{tabular} & $p$-value & \begin{tabular}[c]{@{}l@{}}train OFF \\ test ON\end{tabular} & \begin{tabular}[c]{@{}l@{}}train ON \\ test OFF\end{tabular} & $p$-value \\ \cline{1-7}
accuracy, \% & 55.8$\pm$10.1 & 80.7$\pm$7.1 & $<$ 0.0001* & 55.0$\pm$9.2 & 76.0$\pm$7.2 & $<$ 0.0001* \\ \cline{1-7}
recall, \% & 50.2$\pm$12.2 & 68.1$\pm$12.9 & $<$ 0.0001* & 48.7$\pm$14.1 & 58.6$\pm$13.2 & $<$ 0.0001* \\ \cline{1-7}
specificity, \% & 60.9$\pm$14.1 & 92.4$\pm$7.1 & $<$ 0.0001* & 60.8$\pm$12.7 & 92.1$\pm$6.7 & $<$ 0.0001* \\ \cline{1-7}
precision, \% & 54.8$\pm$12.0 & 89.9$\pm$9.1 & $<$ 0.0001* & 53.6$\pm$10.4 & 87.8$\pm$9.8 & $<$ 0.0001* \\ \cline{1-7}
$F_1$ score, \% & 51.9$\pm$11.1 & 76.7$\pm$9.6 & $<$ 0.0001* & 50.4$\pm$11.2 & 69.4$\pm$10.6 & $<$ 0.0001*\\ \cline{1-7}
AUC & 0.56$\pm$0.10 & 0.80$\pm$0.072 & $<$ 0.0001* & 0.55$\pm$0.092 & 0.75$\pm$0.074 & $<$ 0.0001* \\
\cline{2-3} \cline{5-6}
\end{tabular}
\end{table*}

\begin{table*}[!t]
\centering
\caption{Assessment of generalization: Test results (mean $\pm$ standard deviation (SD)) achieved with a support-vector machine (SVM) model trained and tested on different medication states (ON/OFF). * $p$-values $< 0.05$ were considered statistically significant. Abbreviations: AUC = area under the receiver operating characteristic curve.}
\begin{tabular}{c|c|c|c|c|c|c}
\cline{2-3} \cline{5-6}
& \begin{tabular}[c]{@{}l@{}}train OFF\\ test ON\end{tabular} & \begin{tabular}[c]{@{}l@{}}train ON\\ test ON\end{tabular} & $p$-value & \begin{tabular}[c]{@{}l@{}}train OFF \\ test ON\end{tabular} & \begin{tabular}[c]{@{}l@{}}train ON \\ test OFF\end{tabular} & $p$-value \\ \cline{1-7}
accuracy, \% & 55.8$\pm$10.1 & 55.0$\pm$9.2 & 0.604 & 80.7$\pm$7.1 & 76.0$\pm$7.2 & $<$ 0.0001* \\ \cline{1-7}
recall, \% & 50.2$\pm$12.2 & 48.7$\pm$14.1 & 0.452 & 68.1$\pm$12.9 & 58.6$\pm$13.2 & $<$ 0.0001* \\ \cline{1-7}
specificity, \% & 60.9$\pm$14.1 & 60.8$\pm$12.7 & 0.965 & 92.4$\pm$7.1 & 92.1$\pm$6.7 & 0.759 \\ \cline{1-7}
precision, \% & 54.8$\pm$12.0 & 53.6$\pm$10.4 & 0.440 & 89.9$\pm$9.1 & 87.8$\pm$9.8 & 0.175 \\ \cline{1-7}
$F_1$ score, \% & 51.9$\pm$11.1 & 50.4$\pm$11.2 & 0.366 & 76.7$\pm$9.6 & 69.4$\pm$10.6 & $<$ 0.0001*\\ \cline{1-7}
AUC & 0.56$\pm$0.10 & 0.55$\pm$0.092 & 0.567 & 0.80$\pm$0.072 & 0.75$\pm$0.074 & $<$ 0.0001* \\
\cline{2-3} \cline{5-6}
\end{tabular}
\end{table*}

\subsection{Framework for Parkinson's Disease Detection}
The confounding analysis framework was based on an AI framework for detecting PD. See \cite{kurbatskaya2023_2} for a more detailed description of the ML framework used in this study. Briefly, standard ML models \cite{cleophas2015}, such as logistic regression (LR), decision tree (DT), k-nearest neighbor (k-NN), and support vector machine (SVM) algorithms, were trained and compared based on an unseen cohort during training. The feature data were split 70/30\% for training and testing, stratified by center, gender, and group class labels. This stratification strategy ensured that both training and test sets were equally represented by factors that could potentially introduce bias \cite{raschka2018}. The features of each subject were included in either the training set or the test set, but not both, to prevent cross-contamination. The training set was used for hyperparameter tuning and model selection through a nested stratified 5-fold cross-validation (CV) method, ensuring a robust evaluation of the models. In the inner loop, a grid search algorithm was executed over the parameter space, while in the outer loop, the generalized performance of the resulting algorithms was evaluated. To manage the high dimensionality of the data,  we performed feature selection using univariate feature selection based on ANOVA $F$-values \cite{jovic2015}. We validated the number of features as part of the tunable parameters. The final model was selected based on the highest accuracy averaged over the five CV folds.

\subsection{Statistical Analysis}
To estimate the uncertainty of the performance, the test sets with the final set of features were bootstrapped, resampling with replacement to match the original test set size, over $100$ iterations. The performance of the classifiers was evaluated in terms of accuracy, recall, specificity, precision, $F_{1}$ score, and area under the receiver operating characteristic (ROC) curve (AUC). Each continuous metric is presented as mean$\pm$standard deviation (SD) value. To assess the statistical significance between test results, we performed a paired permutation test over $1000$ permutations. A $p$-value of less than $0.05$ was considered statistically significant.

\section{Results}
The analysis included a total of 40 non-PD individuals and 41 individuals with PD. TABLE 1 summarizes the demographic and clinical characteristics of the two datasets, forming the resulting data cohort. No significant differences were observed between the two groups in terms of age, MMSE scores, or NAART scores across both datasets. The cohort was also gender-balanced, with no significant differences in gender distribution between the groups.

For the final performance evaluation, models were selected based on their validation accuracy. Among the evaluated models, SVM demonstrated the best performance. In the following subsections, we present results from the best-performing models when tested on independent test sets.

\subsection{Assessment of Invariance}

TABLE 2 presents the results of the assessment of invariance. As depicted, in both comparisons, all the metrics were significantly different ($p < 0.001$). Furthermore, the model trained on the ON-medication state consistently outperformed the model trained on the OFF-medication state across all metrics, regardless of whether it was tested on the ON or OFF-medication state.


\subsection{Assessment of Generalization}

TABLE 3 presents the results of the assessment of generalization. As depicted, in both cases, the model performed worse on the alternating medication state than on the matching medication state. However, the model trained on the OFF-medication state showed no significant difference in performance across all metrics when tested on either the OFF- or ON-medication state. In contrast, the model trained on the ON-medication state demonstrated significantly worse performance in terms of accuracy, recall, and $F_1$ score when tested on the OFF-medication state compared to the ON-medication state. Additionally, the model trained on the ON-medication state consistently outperformed the model trained on the OFF-medication state across all metrics, regardless of the test set.

\section{Discussion}
This study examined the impact of medication state as a confounding factor on the training and generalization of EEG-based AI models for detecting PD. Specifically, we focused on assessing how different medication states influence the models' invariance and generalization properties. Our findings highlight that models trained using rs-EEG data acquired under different medication conditions from PD individuals exhibit significant performance differences, which might impact their robustness and clinical applicability in PD management and characterization.

Invariance refers to the model’s ability to maintain consistent performance regardless of variations in irrelevant factors, such as the medication state. Ideally, an invariant model should perform similarly whether the training and test data come from ON-medication or OFF-medication states. In this study, models trained on ON-medication data outperformed those trained on OFF-medication data when tested within the same medication state, achieving $ACC_{\text{ON-ON}}=80.7\pm7.1$ compared to $ACC_{\text{OFF-OFF}}=56.2\pm8.7$ (TABLE 2). This suggests that the ON-trained model may be learning medication-related features that significantly improves its performance in ON-medication data but limits its invariance across medication states.

Generalization, on the other hand, pertains to the model's ability to apply learned patterns effectively to new, unseen data, regardless of variation in medication state (TABLE 3). Our findings showed that models trained on the OFF-medication state demonstrated relatively stable generalization, maintaining performance when tested on both ON and OFF states ($ACC_{\text{OFF-ON}}=55.3\pm8.8$ and $ACC_{\text{OFF-OFF}}=56.2\pm8.7$). Conversely, models trained on ON-medication data experienced a notable drop in generalization when tested on OFF-medication data ($ACC_{\text{ON-OFF}}=76.0\pm7.2$), despite performing well within the ON-medication state. This indicates that the model trained on ON-medication data, while showing high accuracy in ON conditions, fails to generalize effectively to OFF conditions due to over-reliance on medication-related features.

These results highlight that the models trained on ON-medication data are not invariant to medication state and have reduced generalizability when applied to different conditions. The model trained on OFF-medication data, in contrast, exhibits more consistent performance across medication states, suggesting that it may be more robust when applied to diverse real-world scenarios.

The growing need for improved patient management in PD has led to the increased interest of using EEG as an indicator of neural dysfunction, combined with AI. However, the validity of the proposed methods remains questionable due to potential shortcuts, where models rely on confounding factors such as medication state rather than true disease-related patterns. As seen in other subfields of medical AI, models often exacerbate and propagate demographic disparities \cite{brown2023, yang2024}. 

Previous studies have demonstrated the presence of hidden information, such as biological age and gender, in EEG recordings. Our findings suggest that the medication state in rs-EEG is not irrelevant but plays a critical role in model performance. The model’s lack of invariance to medication state indicates that it introduces significant variability, potentially confounding the detection of PD-specific patterns.

These findings highlight the need for models capable of distinguishing between disease-related neural signatures and medication-induced changes in the EEG signals. Without addressing this confounding factor, models may produce biased results that overly depend on the presence or absence of medication, leading to overoptimistic performance and limiting their clinical applicability. Future work should focus on developing strategies to mitigate this confounding effect, ensuring the development of robust, clinically useful models.

\subsection*{Limitations}
Despite merging two publicly available datasets, the number of rs-EEG recordings was relatively low due to the challenges and complexities in acquiring paired rs-EEG recordings from the same PD individuals in different medication states. The retrospective nature of the data also limited our ability to control all the potential confounding factors, which may have influenced the results. Further, more comprehensive analyses are required to validate the consistency of these findings, ideally using larger datasets and prospective data collection to address these limitations and incorporate additional data sources.

\section{Conclusion}
Our findings reveal that the performance of EEG-based AI models for predicting PD-related outcomes is significantly influenced by the medication state of individuals with clinical manifestations of PD. This underscores the necessity of accounting for medication information in the models designed for clinical use in PD management. Ignoring this factor could lead to models that are overly sensitive to medication effects and less reliable in detecting disease-specific patterns.
 
\section*{Acknowledgment}

We would like to thank all the researchers and staff at the University of New Mexico, New Mexico, and the university of California, San Diego, for making their datasets publicly available.

\appendices



\printbibliography[title=References]

@article{zhu2024,
  title={Temporal trends in the prevalence of {P}arkinson's disease from 1980 to 2023: a systematic review and meta-analysis},
  author={Zhu, Jinqiao and Cui, Yusha and Zhang, Junjiao and Yan, Rui and Su, Dongning and Zhao, Dong and Wang, Anxin and Feng, Tao},
  journal={The Lancet Healthy Longevity},
  volume={5},
  number={7},
  pages={e464--e479},
  year={2024},
  publisher={Elsevier}
}

@article{postuma2015,
  title={{MDS} clinical diagnostic criteria for {P}arkinson's disease},
  author={Postuma, Ronald B and Berg, Daniela and Stern, Matthew and Poewe, Werner and Olanow, C Warren and Oertel, Wolfgang and Obeso, Jos{\'e} and Marek, Kenneth and Litvan, Irene and Lang, Anthony E and others},
  journal={Movement disorders},
  volume={30},
  number={12},
  pages={1591--1601},
  year={2015},
  publisher={Wiley Online Library}
}

@article{mulroy2024,
  title={Refining the clinical diagnosis of {P}arkinson's disease},
  author={Mulroy, Eoin and Erro, Roberto and Bhatia, Kailash P and Hallett, Mark},
  journal={Parkinsonism \& Related Disorders},
  pages={106041},
  year={2024},
  publisher={Elsevier}
}

@article{armstrong2020,
  title={Diagnosis and treatment of {P}arkinson disease: a review},
  author={Armstrong, Melissa J and Okun, Michael S},
  journal={Jama},
  volume={323},
  number={6},
  pages={548--560},
  year={2020},
  publisher={American Medical Association}
}

@article{jost2023,
  title={Levodopa dose equivalency in {P}arkinson's disease: updated systematic review and proposals},
  author={Jost, Stefanie T and Kaldenbach, Marie-Ann and Antonini, Angelo and Martinez-Martin, Pablo and Timmermann, Lars and Odin, Per and Katzenschlager, Regina and Borgohain, Rupam and Fasano, Alfonso and Stocchi, Fabrizio and others},
  journal={Movement Disorders},
  volume={38},
  number={7},
  pages={1236--1252},
  year={2023},
  publisher={Wiley Online Library}
}

@article{pahwa2009,
  title={Levodopa-related wearing-off in {P}arkinson's disease: identification and management},
  author={Pahwa, Rajesh and Lyons, Kelly E},
  journal={Current medical research and opinion},
  volume={25},
  number={4},
  pages={841--849},
  year={2009},
  publisher={Taylor \& Francis}
}

@article{bloem2021,
  title={{P}arkinson's disease},
  author={Bloem, Bastiaan R and Okun, Michael S and Klein, Christine},
  journal={The Lancet},
  volume={397},
  number={10291},
  pages={2284--2303},
  year={2021},
  publisher={Elsevier}
}

@article{vanneste2018,
  title={Thalamocortical dysrhythmia detected by machine learning},
  author={Vanneste, Sven and Song, Jae-Jin and De Ridder, Dirk},
  journal={Nature communications},
  volume={9},
  number={1},
  pages={1103},
  year={2018},
  publisher={Nature Publishing Group UK London}
}

@article{betrouni2019,
  title={Electroencephalography-based machine learning for cognitive profiling in {P}arkinson's disease: Preliminary results},
  author={Betrouni, Nacim and Delval, Arnaud and Chaton, Laurence and Defebvre, Luc and Duits, Annelien and Moonen, Anja and Leentjens, Albert FG and Dujardin, Kathy},
  journal={Movement Disorders},
  volume={34},
  number={2},
  pages={210--217},
  year={2019},
  publisher={Wiley Online Library}
}

@article{waninger2020,
  title={Neurophysiological biomarkers of {P}arkinson’s disease},
  author={Waninger, Shani and Berka, Chris and Stevanovic Karic, Marija and Korszen, Stephanie and Mozley, P David and Henchcliffe, Claire and Kang, Yeona and Hesterman, Jacob and Mangoubi, Tomer and Verma, Ajay},
  journal={Journal of Parkinson's disease},
  volume={10},
  number={2},
  pages={471--480},
  year={2020},
  publisher={IOS Press}
}

@article{khoshnevis2021,
  title={Classification of the stages of {P}arkinson’s disease using novel higher-order statistical features of {EEG} signals},
  author={Khoshnevis, Seyed Alireza and Sankar, Ravi},
  journal={Neural Computing and Applications},
  volume={33},
  pages={7615--7627},
  year={2021},
  publisher={Springer}
}

@article{yassine2023,
  title={Identification of {P}arkinson's disease subtypes from resting-state electroencephalography},
  author={Yassine, Sahar and Gschwandtner, Ute and Auffret, Manon and Duprez, Joan and Verin, Marc and Fuhr, Peter and Hassan, Mahmoud},
  journal={Movement Disorders},
  volume={38},
  number={8},
  pages={1451--1460},
  year={2023},
  publisher={Wiley Online Library}
}

@article{sugden2023,
  title={Generalizable electroencephalographic classification of {P}arkinson's disease using deep learning},
  author={Sugden, Richard James and Diamandis, Phedias},
  journal={Informatics in Medicine Unlocked},
  volume={42},
  pages={101352},
  year={2023},
  publisher={Elsevier}
}

@article{suuronen2023,
  title={Budget-based classification of {P}arkinson's disease from resting state {EEG}},
  author={Suuronen, Ilkka and Airola, Antti and Pahikkala, Tapio and Murtoj{\"a}rvi, Mika and Kaasinen, Valtteri and Railo, Henry},
  journal={IEEE Journal of Biomedical and Health Informatics},
  volume={27},
  number={8},
  pages={3740--3747},
  year={2023},
  publisher={IEEE}
}

@article{anjum2024,
  title={Resting-state {EEG} measures cognitive impairment in {P}arkinson’s disease},
  author={Anjum, Md Fahim and Espinoza, Arturo I and Cole, Rachel C and Singh, Arun and May, Patrick and Uc, Ergun Y and Dasgupta, Soura and Narayanan, Nandakumar S},
  journal={npj Parkinson's Disease},
  volume={10},
  number={1},
  pages={6},
  year={2024},
  publisher={Nature Publishing Group UK London}
}

@article{frisoni2024,
  title={European intersocietal recommendations for the biomarker-based diagnosis of neurocognitive disorders},
  author={Frisoni, Giovanni B and Festari, Cristina and Massa, Federico and Ramusino, Matteo Cotta and Orini, Stefania and Aarsland, Dag and Agosta, Federica and Babiloni, Claudio and Borroni, Barbara and Cappa, Stefano F and others},
  journal={The Lancet Neurology},
  volume={23},
  number={3},
  pages={302--312},
  year={2024},
  publisher={Elsevier}
}

@article{engemann2022,
  title={A reusable benchmark of brain-age prediction from {M}/{EEG} resting-state signals},
  author={Engemann, Denis A and Mellot, Apolline and H{\"o}chenberger, Richard and Banville, Hubert and Sabbagh, David and Gemein, Lukas and Ball, Tonio and Gramfort, Alexandre},
  journal={Neuroimage},
  volume={262},
  pages={119521},
  year={2022},
  publisher={Elsevier}
}

@article{khayretdinova2024,
  title={Prediction of brain sex from {EEG}: using large-scale heterogeneous dataset for developing a highly accurate and interpretable {ML} model},
  author={Khayretdinova, Mariam and Zakharov, Ilya and Pshonkovskaya, Polina and Adamovich, Timothy and Kiryasov, Andrey and Zhdanov, Andrey and Shovkun, Alexey},
  journal={NeuroImage},
  volume={285},
  pages={120495},
  year={2024},
  publisher={Elsevier}
}

@inproceedings{kurbatskaya2023_1,
  title={Assessing gender fairness in {EEG}-based machine learning detection of {P}arkinson's disease: A multi-center study},
  author={Kurbatskaya, Anna and Jaramillo-Jimenez, Alberto and Ochoa-Gomez, John Fredy and Br{\o}nnick, Kolbj{\o}rn and Fernandez-Quilez, Alvaro},
  booktitle={2023 31st European Signal Processing Conference (EUSIPCO)},
  pages={1020--1024},
  year={2023},
  organization={IEEE}
}

@article{stige2024,
  title={The {STRAT-PARK} cohort: a personalized initiative to stratify {P}arkinson’s disease},
  author={Stige, Kjersti Eline and Kverneng, Simon Ulvenes and Sharma, Soumya and Skeie, Geir-Olve and Sheard, Erika and S{\o}gnen, Mona and Geijerstam, Solveig Af and Vet{\aa}s, Therese and Wahlv{\aa}g, Anne Grete and Berven, Haakon and others},
  journal={Progress in Neurobiology},
  pages={102603},
  year={2024},
  publisher={Elsevier}
}

@article{babiloni2019,
  title={Levodopa may affect cortical excitability in {P}arkinson's disease patients with cognitive deficits as revealed by reduced activity of cortical sources of resting state electroencephalographic rhythms},
  author={Babiloni, Claudio and Del Percio, Claudio and Lizio, Roberta and Noce, Giuseppe and Lopez, Susanna and Soricelli, Andrea and Ferri, Raffaele and Pascarelli, Maria Teresa and Catania, Valentina and Nobili, Flavio and others},
  journal={Neurobiology of aging},
  volume={73},
  pages={9--20},
  year={2019},
  publisher={Elsevier}
}

@article{jackson2019,
  title={Characteristics of waveform shape in {P}arkinson’s disease detected with scalp electroencephalography},
  author={Jackson, Nicko and Cole, Scott R and Voytek, Bradley and Swann, Nicole C},
  journal={eneuro},
  volume={6},
  number={3},
  year={2019},
  publisher={Society for Neuroscience}
}

@article{anjum2020,
  title={Linear predictive coding distinguishes spectral {EEG} features of {P}arkinson's disease},
  author={Anjum, Md Fahim and Dasgupta, Soura and Mudumbai, Raghuraman and Singh, Arun and Cavanagh, James F and Narayanan, Nandakumar S},
  journal={Parkinsonism \& related disorders},
  volume={79},
  pages={79--85},
  year={2020},
  publisher={Elsevier}
}

@article{khare2021,
  title={Detection of {P}arkinson’s disease using automated tunable {Q} wavelet transform technique with {EEG} signals},
  author={Khare, Smith K and Bajaj, Varun and Acharya, U Rajendra},
  journal={Biocybernetics and Biomedical Engineering},
  volume={41},
  number={2},
  pages={679--689},
  year={2021},
  publisher={Elsevier}
}

@article{barua2021,
  title={Novel automated {PD} detection system using aspirin pattern with {EEG} signals},
  author={Barua, Prabal Datta and Dogan, Sengul and Tuncer, Turker and Baygin, Mehmet and Acharya, U Rajendra},
  journal={Computers in biology and medicine},
  volume={137},
  pages={104841},
  year={2021},
  publisher={Elsevier}
}

@article{aljalal2022,
  title={Detection of {P}arkinson’s disease from {EEG} signals using discrete wavelet transform, different entropy measures, and machine learning techniques},
  author={Aljalal, Majid and Aldosari, Saeed A and Molinas, Marta and AlSharabi, Khalil and Alturki, Fahd A},
  journal={Scientific Reports},
  volume={12},
  number={1},
  pages={22547},
  year={2022},
  publisher={Nature Publishing Group UK London}
}

@dataset{NMdata,
  author = {James F Cavanagh},
  title = {"{EEG}: 3-{S}tim {A}uditory {O}ddball and {R}est in {P}arkinson's"},
  year = {2021},
  doi = {10.18112/openneuro.ds003490.v1.1.0},
  publisher = {OpenNeuro}
}

@dataset{SDdata,
  author = {Alexander P. Rockhill and Nicko Jackson and Jobi George and Adam Aron and Nicole C. Swann},
  title = {"{UC} {S}an {D}iego {R}esting {S}tate {EEG} {D}ata from {P}atients with {P}arkinson's {D}isease"},
  year = {2021},
  doi = {doi:10.18112/openneuro.ds002778.v1.0.5},
  publisher = {OpenNeuro}
}

@article{cavanagh2018,
  title={Diminished {EEG} habituation to novel events effectively classifies {P}arkinson’s patients},
  author={Cavanagh, James F and Kumar, Praveen and Mueller, Andrea A and Richardson, Sarah Pirio and Mueen, Abdullah},
  journal={Clinical Neurophysiology},
  volume={129},
  number={2},
  pages={409--418},
  year={2018},
  publisher={Elsevier}
}

@article{cavanagh2017,
  title={Cognitive states influence dopamine-driven aberrant learning in {P}arkinson's disease},
  author={Cavanagh, James F and Mueller, Andrea A and Brown, Darin R and Janowich, Jacqueline R and Story-Remer, Jacqueline H and Wegele, Ashley and Richardson, Sarah Pirio},
  journal={Cortex},
  volume={90},
  pages={115--124},
  year={2017},
  publisher={Elsevier}
}

@article{george2013,
  title={Dopaminergic therapy in {P}arkinson's disease decreases cortical beta band coherence in the resting state and increases cortical beta band power during executive control},
  author={George, Jobi S and Strunk, Jon and Mak-McCully, Rachel and Houser, Melissa and Poizner, Howard and Aron, Adam R},
  journal={NeuroImage: Clinical},
  volume={3},
  pages={261--270},
  year={2013},
  publisher={Elsevier}
}

@article{keil2014,
  title={Committee report: publication guidelines and recommendations for studies using electroencephalography and magnetoencephalography},
  author={Keil, Andreas and Debener, Stefan and Gratton, Gabriele and Jungh{\"o}fer, Markus and Kappenman, Emily S and Luck, Steven J and Luu, Phan and Miller, Gregory A and Yee, Cindy M},
  journal={Psychophysiology},
  volume={51},
  number={1},
  pages={1--21},
  year={2014},
  publisher={Wiley Online Library}
}

@article{pernet2019,
  title={{EEG-BIDS}, an extension to the brain imaging data structure for electroencephalography},
  author={Pernet, Cyril R and Appelhoff, Stefan and Gorgolewski, Krzysztof J and Flandin, Guillaume and Phillips, Christophe and Delorme, Arnaud and Oostenveld, Robert},
  journal={Scientific data},
  volume={6},
  number={1},
  pages={103},
  year={2019},
  publisher={Nature Publishing Group UK London}
}

@article{appelhoff2019,
  title={{MNE-BIDS}: Organizing electrophysiological data into the {BIDS} format and facilitating their analysis},
  author={Appelhoff, Stefan and Sanderson, Matthew and Brooks, Teon L and van Vliet, Marijn and Quentin, Romain and Holdgraf, Chris and Chaumon, Maximilien and Mikulan, Ezequiel and Tavabi, Kambiz and H{\"o}chenberger, Richard and others},
  journal={Journal of Open Source Software},
  volume={4},
  number={44},
  pages={1896},
  year={2019},
  publisher={Open Journals}
}

@article{gramfort2013,
  title={{MEG} and {EEG} data analysis with {MNE}-Python},
  author={Gramfort, Alexandre and Luessi, Martin and Larson, Eric and Engemann, Denis A and Strohmeier, Daniel and Brodbeck, Christian and Goj, Roman and Jas, Mainak and Brooks, Teon and Parkkonen, Lauri and others},
  journal={Frontiers in Neuroinformatics},
  volume={7},
  pages={267},
  year={2013},
  publisher={Frontiers Media SA}
}

@article{gramfort2014,
  title={{MNE} software for processing {MEG} and {EEG} data},
  author={Gramfort, Alexandre and Luessi, Martin and Larson, Eric and Engemann, Denis A and Strohmeier, Daniel and Brodbeck, Christian and Parkkonen, Lauri and H{\"a}m{\"a}l{\"a}inen, Matti S},
  journal={neuroimage},
  volume={86},
  pages={446--460},
  year={2014},
  publisher={Elsevier}
}

@article{appelhoff2022,
  title={{P}y{PREP}: A Python implementation of the preprocessing pipeline ({PREP}) for {EEG} data},
  author={Appelhoff, S and Hurst, AJ and Lawrence, A and Li, A and Mantilla, R and Yorguin, J and O’Reilly, C and Xiang, L and Dancker, J},
  journal={Zenodo},
  pages={2},
  year={2022}
}

@article{bigdely2015,
  title={The {PREP} pipeline: standardized preprocessing for large-scale {EEG} analysis},
  author={Bigdely-Shamlo, Nima and Mullen, Tim and Kothe, Christian and Su, Kyung-Min and Robbins, Kay A},
  journal={Frontiers in neuroinformatics},
  volume={9},
  pages={16},
  year={2015},
  publisher={Frontiers Media SA}
}

@article{jas2017,
  title={Autoreject: Automated artifact rejection for {MEG} and {EEG} data},
  author={Jas, Mainak and Engemann, Denis A and Bekhti, Yousra and Raimondo, Federico and Gramfort, Alexandre},
  journal={NeuroImage},
  volume={159},
  pages={417--429},
  year={2017},
  publisher={Elsevier}
}

@article{lee1999,
  title={Independent component analysis using an extended infomax algorithm for mixed subgaussian and supergaussian sources},
  author={Lee, Te-Won and Girolami, Mark and Sejnowski, Terrence J},
  journal={Neural computation},
  volume={11},
  number={2},
  pages={417--441},
  year={1999},
  publisher={MIT Press}
}

@article{li2022,
  title={{MNE}-{ICAL}abel: Automatically annotating {ICA} components with {ICL}abel in {P}ython},
  author={Li, Adam and Feitelberg, Jacob and Saini, Anand Prakash and H{\"o}chenberger, Richard and Scheltienne, Mathieu},
  journal={Journal of Open Source Software},
  volume={7},
  number={76},
  pages={4484},
  year={2022}
}

@article{slepian1961,
  title={Prolate spheroidal wave functions, {F}ourier analysis and uncertainty—{I}},
  author={Slepian, David and Pollak, Henry O},
  journal={Bell System Technical Journal},
  volume={40},
  number={1},
  pages={43--63},
  year={1961},
  publisher={Wiley Online Library}
}

@article{babiloni2020,
  title={International {F}ederation of {C}linical {N}europhysiology ({IFCN})--{EEG} research workgroup: Recommendations on frequency and topographic analysis of resting state {EEG} rhythms. {P}art 1: Applications in clinical research studies},
  author={Babiloni, Claudio and Barry, Robert J and Ba{\c{s}}ar, Erol and Blinowska, Katarzyna J and Cichocki, Andrzej and Drinkenburg, Wilhelmus HIM and Klimesch, Wolfgang and Knight, Robert T and da Silva, Fernando Lopes and Nunez, Paul and others},
  journal={Clinical Neurophysiology},
  volume={131},
  number={1},
  pages={285--307},
  year={2020},
  publisher={Elsevier}
}

@article{vallat2021,
  title={An open-source, high-performance tool for automated sleep staging},
  author={Vallat, Raphael and Walker, Matthew P},
  journal={Elife},
  volume={10},
  pages={e70092},
  year={2021},
  publisher={eLife Sciences Publications Limited}
}

@inproceedings{kurbatskaya2023_2,
  title={Machine learning-based detection of {P}arkinson’s disease from resting-state {EEG}: A multi-center study},
  author={Kurbatskaya, Anna and Jaramillo-Jimenez, Alberto and Ochoa-Gomez, John Fredy and Br{\o}nnick, Kolbj{\o}rn and Fernandez-Quilez, Alvaro},
  booktitle={2023 45th Annual International Conference of the IEEE Engineering in Medicine \& Biology Society (EMBC)},
  pages={1--4},
  year={2023},
  organization={IEEE}
}

@book{cleophas2015,
  title={Machine learning in medicine-a complete overview},
  author={Cleophas, Ton J and Zwinderman, Aeilko H and others},
  volume={21},
  year={2015},
  publisher={Springer}
}

@article{raschka2018,
  title={Model evaluation, model selection, and algorithm selection in machine learning},
  author={Raschka, Sebastian},
  journal={arXiv preprint arXiv:1811.12808},
  year={2018}
}

@inproceedings{jovic2015,
  title={A review of feature selection methods with applications},
  author={Jovi{\'c}, Alan and Brki{\'c}, Karla and Bogunovi{\'c}, Nikola},
  booktitle={2015 38th {I}nternational convention on information and communication technology, electronics and microelectronics ({MIPRO})},
  pages={1200--1205},
  year={2015},
  organization={{IEEE}}
}

@article{brown2023,
  title={Detecting shortcut learning for fair medical {AI} using shortcut testing},
  author={Brown, Alexander and Tomasev, Nenad and Freyberg, Jan and Liu, Yuan and Karthikesalingam, Alan and Schrouff, Jessica},
  journal={Nature communications},
  volume={14},
  number={1},
  pages={4314},
  year={2023},
  publisher={Nature Publishing Group UK London}
}

@article{yang2024,
  title={The limits of fair medical imaging {AI} in real-world generalization},
  author={Yang, Yuzhe and Zhang, Haoran and Gichoya, Judy W and Katabi, Dina and Ghassemi, Marzyeh},
  journal={Nature Medicine},
  pages={1--11},
  year={2024},
  publisher={Nature Publishing Group US New York}
}

\EOD
\end{document}